\documentclass[prb,twocolumn,superscriptaddress,showpacs]{revtex4}
\usepackage[latin1]{inputenc}
\usepackage{amsmath}
\usepackage{amsfonts}
\usepackage{amssymb}
\usepackage{subfigure}
\usepackage{graphicx}
\usepackage{epstopdf}
\usepackage{epsfig}

\begin{document}

\addtolength{\topmargin}{0.125in}
\addtolength{\oddsidemargin}{0.00in}

\pagestyle{empty}

\title{The gradual destruction of magnetism in the superconducting family NaFe$_{1-x}$Co$_x$As}
\author{J.~D.~Wright}
\affiliation{Department of Physics, Clarendon Laboratory, University of Oxford, Parks Road, Oxford, OX1 3PU, United Kingdom }

\author{T.~Lancaster}
\affiliation{Department of Physics, Clarendon Laboratory, University of Oxford, Parks Road, Oxford, OX1 3PU, United Kingdom }

\author{I.~Franke}
\affiliation{Department of Physics, Clarendon Laboratory, University of Oxford, Parks Road, Oxford, OX1 3PU, United Kingdom }

\author{A.~J.~Steele}
\affiliation{Department of Physics, Clarendon Laboratory, University of Oxford, Parks Road, Oxford, OX1 3PU, United Kingdom }

\author{J.~S.~M\"oller}
\affiliation{Department of Physics, Clarendon Laboratory, University of Oxford, Parks Road, Oxford, OX1 3PU, United Kingdom }

\author{M.~J.~Pitcher}
\affiliation{Department of Chemistry, University of Oxford, South Parks Road, Oxford, OX1 3QR, United Kingdom }

\author{A.~J.~Corkett}
\affiliation{Department of Chemistry, University of Oxford, South Parks Road, Oxford, OX1 3QR, United Kingdom }

\author{D.~R.~Parker}
\affiliation{Department of Chemistry, University of Oxford, South Parks Road, Oxford, OX1 3QR, United Kingdom }

\author{D.~G.~Free}
\affiliation{Department of Chemistry, University of Oxford, South Parks Road, Oxford, OX1 3QR, United Kingdom }

\author{F.~L.~Pratt}
\affiliation{ISIS Muon Facility, ISIS, Chilton, Oxon, OX11 0QX, United Kingdom }

\author{P.~J.~Baker}
\affiliation{ISIS Muon Facility, ISIS, Chilton, Oxon, OX11 0QX, United Kingdom }

\author{S.~J.~Clarke}
\affiliation{Department of Chemistry, University of Oxford, South Parks Road, Oxford, OX1 3QR, United Kingdom }

\author{S.~J.~Blundell}
\affiliation{Department of Physics, Clarendon Laboratory, University of Oxford, Parks Road, Oxford, OX1 3PU, United Kingdom }

\begin{abstract}
The interplay and coexistence of superconducting, magnetic and
structural order parameters in NaFe$_{1-x}$Co$_{x}$As has been studied
using SQUID magnetometry, muon-spin rotation and synchrotron x-ray powder diffraction. Substituting Fe by Co weakens the ordered magnetic state through both a
suppression of $T_{\mathrm{N}}$ and a reduction in the size of the ordered moment. Upon further substitution of Fe by Co the high sensitivity of the muon as a local magnetic probe reveals a
magnetically disordered phase, in which the size of the moment
continues to decrease and falls to zero around the same point at which the magnetically-driven
structural distortion is no longer resolvable. Both magnetism and the structural distortion are weakened
as the robust superconducting state is established. 
\end{abstract}  


\maketitle

\section{Introduction}
The phase diagrams of iron-based superconductors demonstrate that, as
for the cuprates, the superconducting state generally emerges from a
magnetic parent compound under chemical substitution
\cite{Hosono_Rev,Norman_Rev,Nandi122}. However, in contrast to the
cuprates where the magnetic state is due to electrons localized in a
Mott insulator, in iron-based superconductors it arises from an
instability of the delocalised Fe d-band electrons \cite{Mazin1}. It
is important to understand how this magnetic state evolves
into the superconducting one and how the two states compete, coexist
or mutually exclude one another.

\begin{figure}[h]
\includegraphics[width=\columnwidth, trim=1cm 1cm 2cm 0cm]{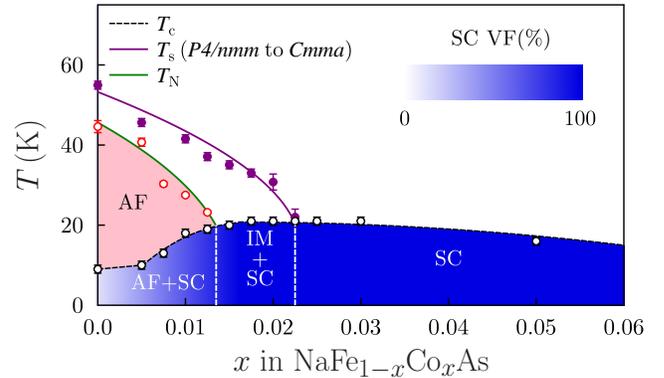}
\caption{Phase diagram for NaFe$_{1-x}$Co$_{x}$As compiled using data from
  XRPD ($T_{\rm s}$), ZF-$\mu$SR ($T_{\rm N}$) and SQUID magnetometry
  ($T_{\rm c}$) and showing regions of antiferromagnetism (AF),
  superconductivity (SC) and inhomogeneous magnetism (IM), as well
  as their regions of coexistence. The shading of the SC region corresponds to the volume fraction estimated from susceptibility data.\label{phaseplot}}
\end{figure}

In this paper we study the ``111" arsenide compound NaFe$_{1-x}$Co$_{x}$As using SQUID magnetometry, muon spin rotation ($\mu$SR) and
synchrotron x-ray powder diffraction (XRPD) experiments. Our results allow us to
follow in detail how the magnetic state collapses as Fe is substituted by Co, first
by weakening the magnetic long-range order (LRO) and structural
distortion and then by suppressing magnetic fluctuations. Our data
demonstrate the decisive influence of the magnetism on the system and
show how its disappearance correlates with the strengthening of the
superconducting phase.  A summary of our findings is shown in the
phase diagram in Fig.~\ref{phaseplot} (which includes some data points
from Ref.~\onlinecite{NaFeAs_prl10}).  It is notable that, in
comparison with other pnictides, very small amounts of cobalt on the Fe site
result in a complete suppression of the magnetic state. For example,
the optimal $T_{\rm c}$ is obtained with a substitution of $\sim
1.5\%$ Co on the Fe site \cite{NaFeAs_prl10} in NaFe$_{1-x}$Co$_x$As,
compared to \cite{Nandi122} $\sim 6.5\%$ Co in Ba(Fe$_{1-x}$Co$_x$)$_2$As$_2$. This means that significantly smaller structural and
electronic changes are being made to the stoichiometric system thereby
reducing any effect of inhomogeneous substitution on the properties under investigation. 

For this study, a series of NaFe$_{1-x}$Co$_x$As compounds were synthesised according to the methods described in Ref.~\onlinecite{NaFeAs_prl10} and details of structural characterisation and purity checks can be found there.

\section{SQUID Magnetometry}

\begin{figure}[h]
\includegraphics[width=\columnwidth]{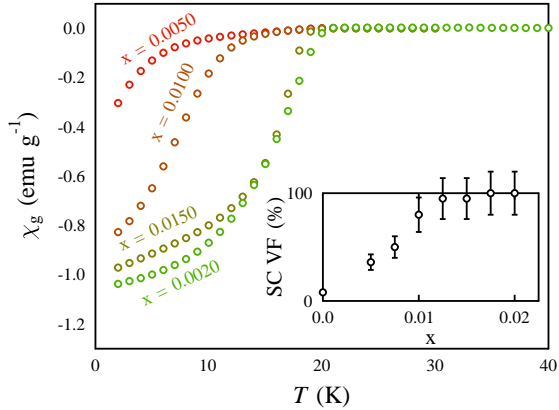}
\caption{Susceptibility data for various NaFe$_{1-x}$Co$_{x}$As samples. {\it Inset}: estimated volume fractions for all samples with $0 \leq x\leq0.02 $. A full superconducting volume fraction is established around $x \approx 0.0125$.\label{SQUID}}
\end{figure}

Magnetic susceptibility measurements were carried out on a Quantum Design MPMS-XL SQUID magnetometer under zero-field cooled and field cooled conditions in a measuring field of $5~{\rm mT}$. The evolution of zero-field susceptibility with temperature for a representative set of samples is shown in Fig.~\ref{SQUID} (a more complete data set can be found in Ref.~\onlinecite{NaFeAs_prl10} and the superconducting $T_{\rm c}$ values for all studied samples are indicated in Fig. \ref{phaseplot}). Also shown is the evolution of the estimated superconducting volume fraction with doping [Fig.~\ref{SQUID}\,(inset), shown also by the shading of the superconducting phase in Fig. \ref{phaseplot}], which suggests that samples with $x > 0.0125$ are fully superconducting. No traces of magnetic impurities, such as Fe, were identified by these measurements in any of our samples.     

\section{Zero-field $\mu$SR}

\begin{figure}[h]
\includegraphics[width=\columnwidth, trim=0cm 3cm 0cm 3cm]{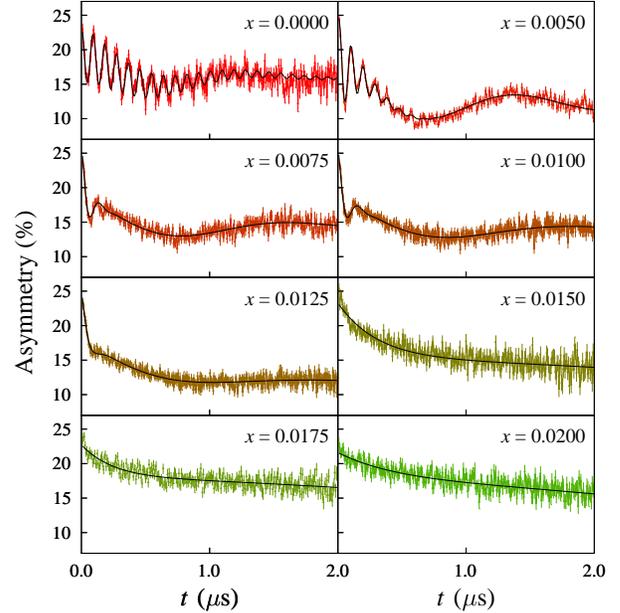}
\caption{Comparison of the ZF $\mu$SR asymmetry spectra for NaFe$_{1-x}$Co$_x$As, 
measured around $1.5~{\rm K}$. \label{rbc}}
\end{figure}

To probe the magnetic order parameter in NaFe$_{1-x}$Co$_x$As,
zero-field (ZF) $\mu$SR measurements were made on samples with $0 \leq
x \leq 0.02$ using the GPS spectrometer at the Swiss Muon
Source and the MuSR spectrometer at the ISIS facility. 
Fig.~\ref{rbc} shows example spectra measured at $T=1.5$~K.  For small
$x$, we observe oscillations in the time dependence of the muon spin
polarization (via the positron decay asymmetry), which are strongly
indicative of magnetic long-range order (LRO).  These become less distinct with
increasing $x$ and are not discernible for $x > 0.0125$.  The
frequencies of the oscillations are proportional to the local magnetic
field at the muon site(s) and scale with the ordered moment, so any one of them may be
considered an effective magnetic order parameter for the system. 

To extract the order parameter as a function of temperature and locate
$T_{\mathrm{N}}$, the oscillating asymmetry function for the materials with $x \leq 0.0125$ were fitted to the expression
$A(t)=\sum_{i=1}^n{A_i}\cos(2\pi\nu_it)\exp(-\lambda_{i}t)$, with the frequencies
held in fixed proportion. For all samples, except stoichiometric NaFeAs, it was found that two frequencies were required to fit the data (i.e. $n=2$) with one around 20 times larger than the other. In the undoped compound the addition of a third component with a frequency very similar in value to the the other low frequency component, slightly improved the fit \cite{NaFeAs_prl10}. As the damping rates, $\lambda_{i}$, of all oscillations increase with $x$, suggesting a rising level of magnetic disorder, we suggest that the third frequency is not resolvable in the doped compounds. For all samples with $0 \leq x \leq 0.0125$, it was found that the amplitude of the high frequency component, $A_1$, made up between 55-65$\%$ of the total oscillating asymmetry and the low frequency component(s) accounted for the remainder. Muon precession frequencies for this series, extrapolated to zero-temperature, are shown in Table \ref{muon_freq} along with those observed in other iron-arsenide systems for comparison. It is notable that the high frequency component in NaFeAs is markedly lower than that observed in the other iron-arsenides listed, and we return to this point later.

\begin{table}[h]
\caption{Muon precession frequencies extrapolated to $T=0$~K, along with amplitudes, for various FeAs-based compounds (* indicates majority component). \label{muon_freq}}
\begin{tabular*}{\columnwidth}{@{\extracolsep{\fill}}c c c}
\hline \hline
Compound & Frequencies (MHz) & Ref. \\ 
\hline NaFeAs & 10.9*, 0.9, 0.5 & Our Work \\ 
\\
LaFeAsO & 23*, 3 & [\onlinecite{KlaussLaFeAsO}] \\
SrFeAsF & 22.2* , 2.0 & [\onlinecite{BakerSrFeAsF}] \\
\\
BaFe$_2$As$_2$ & 28.8*, 7 & [\onlinecite{AczelBaFe2As2}] \\ 
SrFe$_2$As$_2$ & 44* , 13 & [\onlinecite{JescheSrFe2As2}] \\
\\
FeAs & 38.2, 22.7* & [\onlinecite{baker-impure}] \\
\hline \hline 
\end{tabular*}
\end{table} 

For the remaining samples ($0.015 \leq x \leq 0.02$) a fast-relaxing component is observed at early times that disappears as
$x$ increases. These data are best fitted to a relaxation function
$A(t)=A_{\mathrm{slow}}\exp(-\sigma^2 t^2)+A_{\rm fast}\exp(-\lambda_{\rm
  f}t)$, which combines a slowly-relaxing 
Gaussian function, attributable to the magnetism of static and
disordered moments probably of nuclear origin (a Gaussian Kubo-Toyabe function with the same magnetic field distribution fits just as well
in this regime), added to a fast exponential
relaxation.  This fast relaxing component is most likely attributable
to magnetic fluctuations of electron spins which quickly dephase
muons.  The fact that these two contributions are summed
suggests that only a fraction of the muons are dephased by the
electronic fluctuations. This implies that, for $x > 0.0125$, samples show some degree of inhomogeneous magnetism (IM).  For the
materials in this region we observed a decrease in the relative amplitude
of the fast component, $A_{\rm fast}$, indicating that the
extent of the magnetically fluctuating regions decreases as $x$ increases. Our muon data show no sign of signals from the most common impurities in iron arsenide compounds \cite{baker-impure}.

\begin{figure}[h]
\includegraphics[width=\columnwidth, trim=0cm 1cm 0cm 0cm]{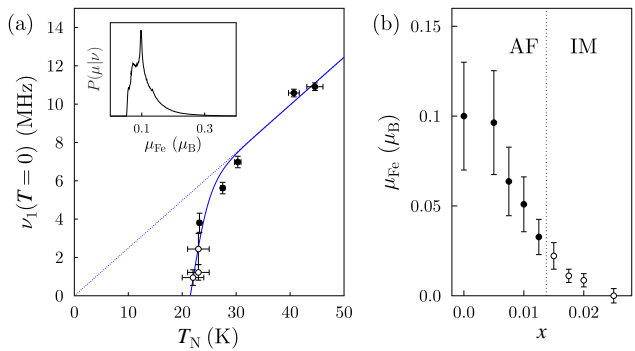}
\caption{(a) Precession frequency (proportional to ordered moment) as
  a function of N\'eel temperature shown by filled circles. In the IM
  state, the effective $\nu_1(0)$ is estimated from the magnetic
  contribution to the rms field width and $T_{\rm N}$ is taken to be the temperature at which this contribution sets in [see Fig. \ref{AllTF}]. {\it Inset:} The probability distribution for the ordered moment for
stoichiometric NaFeAs given the observed precession frequency.
(b) The estimated ordered moment as a function of $x$ in the AF state
(filled circles) and IM state (open circles). \label{As2}}
\end{figure}

The low values of the ZF muon precession frequencies in NaFeAs,
compared to those seen in other iron-based superconductors (see Table \ref{muon_freq}), indicate that the moment is correspondingly low.
Neutron scattering measurements indeed indicate a low moment,
0.09(4)\,$\mu_{\rm B}$, on the Fe site in pure \cite{Li_NaFeAs_moment} NaFeAs. Our ZF-$\mu$SR data can be used to make an
independent estimate of this moment, along with its variation with $x$.  We can obtain the probability $P(\mu_{\rm Fe}|\lbrace
\nu_i\rbrace)$ [shown in the inset of Fig.~\ref{As2}(a)], that is the
probability of $\mu_{\rm Fe}$ taking a particular value given that we
observe a set of precession frequencies, $\lbrace \nu_i\rbrace$, using
Bayes' theorem to invert the probability of observing a set of
precession frequencies given a particular Fe moment (which is easily
calculable from dipole-field simulations). This method (see
Refs.~\onlinecite{Steele_molecular} and ~\onlinecite{SJB_Bayesian}) allows us to provide a quantitative estimate
of $\mu_{\rm Fe}=0.10(3)\mu_B$, assuming the magnetic structure to be as
determined in Ref.~\onlinecite{Li_NaFeAs_moment}, ignoring any
hyperfine coupling and constraining the muon location to be at least $1\,{\rm \AA}$ from an atom site. This estimate is in good agreement with the value from neutron scattering \cite{Li_NaFeAs_moment}. The dependence of $\nu_{1}(0)$ as a
function of $T_{\rm N}$ is shown by the solid circles in
Fig.~\ref{As2}(a) and hence the dependence of $\mu_{\rm Fe}$ as a
function of  $x$ is shown in Fig.~\ref{As2}(b). It is
noticeable that $\nu_{1}(0)$ (and hence $\mu_{\rm Fe}$) tends to
zero more quickly than $T_{\rm N}$. This suggests it is the collapse of the moment that destroys the magnetic state.

\section{X-ray Powder Diffraction (XRPD)}

\begin{figure}[h]
\includegraphics[width=\columnwidth, trim=0cm 0cm 0cm 2cm]{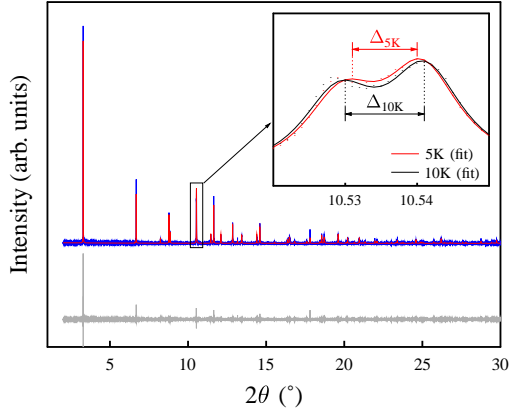}
\caption{A Pawley refinement for NaFe$_{0.98}$Co$_{0.02}$As at $5\,{\rm K}$. {\it Inset:} A comparison of the fits to the 022 and 202 peaks in $Cmma$ at $5\,{\rm K}$ and $10\,{\rm K}$. These peaks converge into the 112 peak in $P4/nmm$. Note that $\Delta$, the splitting in the $2\theta$ values of the two peaks, is smaller at $5\,{\rm K}$ than at $10\,{\rm K}$, suggesting a suppression of the structural distortion at low temperatures. This is in agreement with the behavior of $\delta(T)$ as obtained from fitting the full pattern. \label{peakview}}
\end{figure}
\begin{figure}[h]
\includegraphics[width=\columnwidth, trim=0cm 0cm 0cm
  0cm]{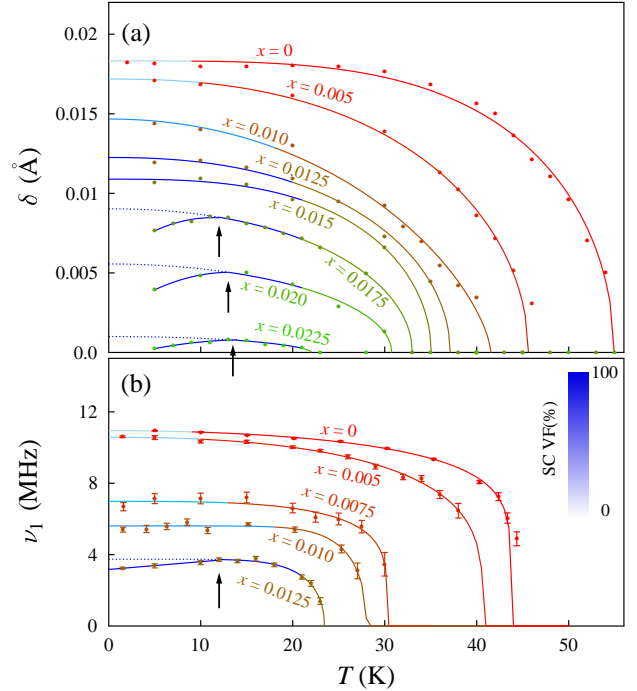}
\caption{The evolution of (a) structural and (b) magnetic order
  parameters with $T$ and $x$. $\delta$ is defined as the difference between the $a$ and $b$ lattice parameters in the low-temperature $Cmma$ phase and $\nu_1$ is the largest observed muon precession frequency in zero-field. The blue line sections show
  where the sample superconducts and the shade of blue indicates the
  volume fraction obtained from SQUID
  magnetometry. The arrows indicate the suppression of $\delta$ and $\nu_1$ in the more highly substituted samples. \label{struct_muon}}
\end{figure}

To probe the structural distortion in NaFe$_{1-x}$Co$_x$As we performed XRPD studies of samples with $0<x<0.025$ on the ID31 beamline at ESRF. In Ref. \onlinecite{NaFeAs_prl10} the distortion was characterised as a transition from a high temperature tetragonal ($P4/nmm$) phase to a low temperature orthorhombic ($Cmma$) phase, with $a_{\rm orth} = \sqrt{2}a_{\rm tet}$. In order to analyse the variation in the size of this distortion with temperature and doping, we fitted data at all temperatures to a $Cmma$ phase. In the TOPAS Academic software \cite{TOPAS}, the lattice parameters were defined as $a = \sqrt{2}a_{\rm tet} + \delta/2$ and $b = \sqrt{2}a_{\rm tet} - \delta/2$, with both $a_{\rm tet}$ and $\delta$ allowed to vary. The data were then fitted to this model using a structure-independent Pawley refinement \cite{Pawley} and a typical refinement is shown in Fig. \ref{peakview}. At high temperatures and/or dopings, where $\delta$ was sufficiently low as to only cause peaks to broaden rather than split, a set of refinements with $\delta$ fixed at zero (i.e. using a tetragonal model) was carried out. Comparing the statistics of the two models allowed us to confirm the temperatures at which the structural distortion was no longer resolvable. Plots of $\delta$ vs T for all samples with a resolvable distortion are shown in Fig.~\ref{struct_muon}(a). These data sets have been fitted to the function $\delta=[1-(T/T_{\rm s})^{\alpha}]^{\beta}$ to locate $T_{\rm s}$, with $\alpha$ allowed to vary and $\beta$ fixed around the mean-field value of $1/2$.

Combining $T_{\mathrm{N}}$ from $\mu$SR, $T_{\mathrm{s}}$ from XRPD and
measurements of superconducting $T_{\mathrm{c}}$ and volume fraction
from SQUID magnetometry, allows us to produce the phase diagram in Fig.~\ref{phaseplot}, showing that the
superconductivity, magnetism and structural distortion coexist for
$x\leq 0.0125$, with the superconducting volume fraction increasing
with increasing $x$.  The results of our ZF-$\mu$SR experiments indicate
that the magnetism weakens as $x$ is increased, with the system unable
to sustain magnetic LRO for $x > 0.0125$. However, IM and the structural distortion coexist with bulk superconductivity in the region $0.015\leq x\leq0.020$.

Further insight into the effect of the reduction in the strength of
the magnetism may be obtained by comparing the magnetic and structural
order parameters, $\nu_{1}(T)$ and $\delta(T)$ respectively, as shown
in Fig.~\ref{struct_muon}. A reduction in both $\nu_1(0)$ and $\delta(0)$ is
apparent as $x$ is increased, along with a marked suppression of these
order parameters with decreasing $T$ in the more highly doped
samples (indicated by arrows in Fig. \ref{struct_muon}). The suppression of the structural distortion is evident in Fig.~\ref{peakview}({\it inset}). In this figure the difference in the fitted $2\theta$ values ($\Delta$) of the orthorhombic 022 and 202 reflections (into which the tetragonal 122 peak splits) falls on cooling from $10~{\rm K}$ to $5~{\rm K}$ (having increased on cooling from $T_{\mathrm{s}}$ to $10~{\rm K}$). We note that the suppression of both $\delta$ and $\nu$ at low
$T$ is observed in those samples with a full superconducting volume
fraction, in agreement with what is seen in other pnictide families
\cite{Nandi122, DKPratt_magSC}.

\section{Transverse-field $\mu$SR}

\begin{figure}[h]
\includegraphics[width=\columnwidth, trim=0cm 3cm 0cm 2cm]{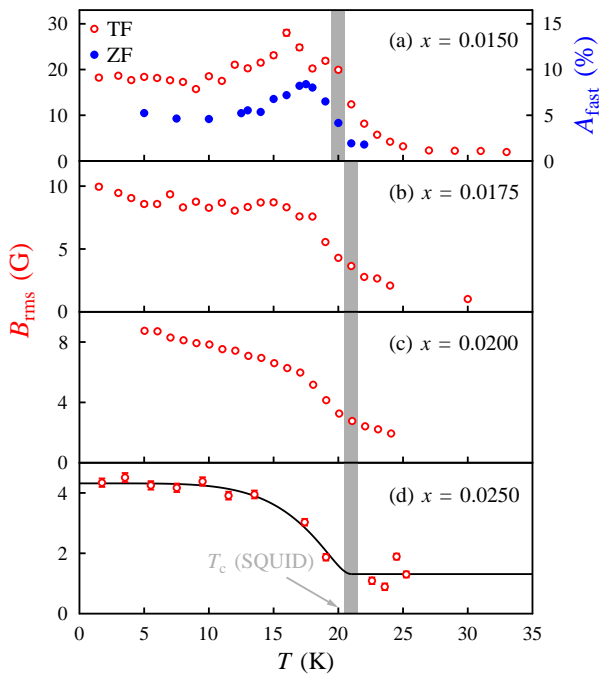}
\caption{The width $B_{\rm rms}$ vs $T$ for samples close to the
  magnetic LRO region.  For $x=0.0150$ (a) the unusual peak in $B_{\rm
    rms}$ matches one observed in the zero-field fast relaxation amplitude, $A_{\rm fast}$.  The
  grey line indicates $T_{\rm c}$ from SQUID magnetometry (see
  Fig.~\ref{phaseplot}). The black line in (d) is a fit to a conventional SC order parameter temperature dependence. \label{AllTF}}
\end{figure}

To further probe the region above $x=0.015$ we have used transverse
field $\mu$SR, which also provides a method of probing the
superconducting state of type-II superconductors
\cite{SCMuSRrevSonier}.  A magnetic field is applied perpendicular to the
initial muon-spin direction, which also produces a vortex lattice in the material's
superconducting phase.  This results in a distribution in the local
field across the sample, whose rms width $B_{\rm rms}$ can be extracted from the muon
precession.  Contributions to $B_{\rm rms}$ arise from
superconductivity, static electronic magnetism and nuclear moments
(the latter being small and temperature-independent) and add in
quadrature. The measured temperature dependence of $B_{\rm rms}$ is
plotted in Fig.~\ref{AllTF} and for each sample studied we find that
$B_{\rm rms}$ rises sharply on cooling through the superconducting $T_{\rm c}$
(found independently from SQUID magnetometry) consistent with a
contribution to the broadening from the vortex lattice.  In addition,
we observe a contribution to $B_{\rm rms}$ from magnetism which
decreases steadily as $x$ increases from 0.015.  For $x=0.025$ the
magnetic contribution is negligible and the temperature dependence of
$B_{\rm rms}$ is consistent with that expected for conventional
superconducting order [solid line in Fig.~\ref{AllTF}(d)].
It is likely that in the materials for $0.015 \leq x < 0.025$ the
IM is associated with a non-zero moment on the
Fe site, but the moment is insufficient to stabilize a fully ordered
magnetic state.   We then associate the increase of $B_{\rm rms}$
just above the superconducting $T_{\rm c}$ with the onset of the IM state.
We also observe a peak in $B_{\rm rms}$ at around 17\,K (most noticeable
in the sample with $x=0.015$ but weakened and broadened for
higher $x$) which may be due to increased correlation between the static moments
(but which stops short of LRO, as we do not observe a precession signal
in ZF-$\mu$SR). This peak is also observed in the ZF amplitude of the fast-relaxing component for the $x=0.015$ sample [Fig. \ref{AllTF}(a)], lending further weight to this interpretation.


We can extract an estimate of the magnetic contribution
to $B_{\rm rms}$ in the IM regime by assuming that a magnetic contribution
and a non-magnetic contribution (the latter assumed to be that of the
$x=0.025$ sample) add in quadrature.  This rms field width
can then be used to provide an estimate
of the precession frequency [$(\gamma_\mu/2\pi) B_{\rm rms}$]
that would have been
observed if the moments in the IM state ordered, and we
plot these as open circles in Fig.~\ref{As2}(a).  By assuming the same scaling between
frequency and $\mu_{\rm Fe}$ as for the AF-ordered region, we can use these values to estimate
$\mu_{\rm Fe}$ which is plotted in Fig.~\ref{As2}(b) using open circles. These data show that as $x$ increases,
the moment on the Fe site undergoes a process of steady
suppression, initially resulting in a loss of AF order as
the order loses long-range coherence but then collapsing
further in the IM state.

\section{Discussion}

It has recently been suggested that the structural transition
in the iron pnictides is driven by magnetic fluctuations \cite{FernandesNematic}, rather than by magnetic order.
The fluctuations could be of different types: spin-nematic fluctuations \cite{FernandesNematic,122nematic}, ferro-orbital fluctuations
or critical fluctuations \cite{cano2010,paul2011}.  Our discussion will focus on the spin-nematic model
of Ref.~\onlinecite{FernandesNematic} which we believe explains the main features
of our data. The theoretical picture is motivated by the
magnetic structure found in these materials, where Fe
spins are coupled antiferromagnetically for Fe-Fe bonds along one diagonal
of the two-Fe tetragonal unit cell and
ferromagnetically along the other diagonal. This
results in two coupled antiferromagnetic sublattices with coplanar staggered magnetization.
The sublattice coupling is related to a
$Z_{2}$ symmetry, which allows two
possible orientations of stripes in the magnetic structure. This symmetry is broken by
{\it nematic order}, which can be present with zero sublattice magnetization and
whose onset may therefore occur at temperatures above any magnetic ordering temperature.
The result of this nematic order is to bias the magnetic fluctuations along a
particular stripe orientation.
Crucially, the nematic order parameter couples to the lattice, causing bonds
between neighboring (anti)parallel spins to (expand) contract and this leads to the
observed structural distortion. The prerequisite for nematic order is
a large, but finite, magnetic correlation length, whose
occurrence at elevated temperatures is probable for materials
in both the magnetically ordered and IM region of the
phase diagram of NaFe$_{1-x}$Co$_{x}$As.

From this point of view
it is magnetic fluctuations, biased by nematic order,
that not only lead to magnetic order, but also drive the structural distortion.
Substituting Fe by Co causes a reduction in the strength of these magnetic
fluctuations and this suppresses both the magnetic order and the structural distortion.
Our measurements show that on increasing $x$ the structural and
magnetic order parameters weaken until the fluctuations
can no longer sustain magnetic order above $x\gtrsim 0.0125$.
Further substitution of Fe by Co continues to weaken the
fluctuations and this coincides with the
weakening of the structural distortion order parameter until fluctuations
can no longer sustain the structural order parameter for $x>0.0225$.
This weakening of the magnetic fluctuations
is coincident with a strengthening of superconductivity in NaFe$_{1-x}$Co$_{x}$As.
This is also consistent with the nematic order picture, which
predicts a competition between superconductivity and magnetism.
Specifically, the onset of superconductivity is predicted to lead to a
reduction in the static part of the
magnetic susceptibility \cite{FernandesNematic} which weakens
the magnetic order and spin correlations. This is seen in our data,
where we observe the low $T$ suppression of
(i) the magnetic order parameter in the $x= 0.0125$ material
(where long-range magnetic order is only just sustained)
and (ii) the structural
order parameter for $x> 0.15$
(where the structural order parameter has been weakened).
We note that these samples have
close to full
superconducting volume fraction and the dip in the order parameter is
only observed for $T\ll T_{\mathrm{c}}$,
where the superconducting order parameter has become sufficiently
strong.

It is interesting to compare these results with those obtained on another ``111" superconductor,
LiFeAs \cite{pitcher2008,wang2008,tapp2008}.
A result of accommodating the smaller Li$^{+}$ ion (rather than Na$^+$) between the FeAs layers
is that the edge-staring FeAs$_4$ tetrahedra in LiFeAs are very compressed
in the basal plane. Superconductivity occurs in
the stoichiometric material LiFeAs, in contrast to other iron
arsenide superconductors for which doping away from the formal oxidation
state of Fe$^{2+}$ or the application of hydrostatic pressure is required
to induce superconductivity. Furthermore, stoichiometric LiFeAs does not appear to show
static magnetism \cite{Taylor_LiFeAs} (in contrast to NaFeAs).
The present study demonstrates that substituing a small amount of Co on the Fe site strengthens the superconducting state in NaFeAs,
but in LiFeAs substitution of Fe by small amounts of Co or Ni results in a steady
lowering of the superconducting transition temperature \cite{pitcher2010}.  For LiFeAs, $T_{\rm c}$
is lowered monotonically at a rate of 10\,K per 0.1 electrons added per formula unit
irrespective of whether the dopant is Co and Ni, and at higher substitution levels superconductivity is completely
suppressed.  (Co and Ni have a similar {\it structural} effect as a function of the level of doping, but Ni adds twice as many electrons as Co.)
The number of electrons added per formula unit is also the determinative quantity for NaFeAs, but here the superconducting state is first strengthened (as the magnetic state is destroyed) and then weakened (see Fig.~1).

Finally, the behavior of the superfluid stiffness as a function of $T_{\rm c}$ in LiFeAs derivatives is markedly different from that of
other pnictides \cite{pratt2009}, including the isostructural NaFeAs-derivatives considered here
and the ``1111" and ``122" classes \cite{pitcher2010}. Most of the pnictides exhibit
behavior more similar to the hole-doped cuprates, but LiFeAs derivatives
resemble much more closely the electron-doped cuprates.
The differing behavior of LiFeAs may result from the underlying structural difference introduced by the small Li$^+$ ion, resulting
in a band structure which does not favor Fermi
surface nesting \cite{borisenko2010} so that
the magnetic instability found in the other isoelectronic
(i.e. undoped) iron arsenides does not compete successfully
against superconductivity in LiFeAs.

\section{Conclusion}

In conclusion, we have presented a picture of NaFe$_{1-x}$Co$_{x}$As
in which a magnetic interaction drives both magnetic
LRO and a structural distortion and weakens with the subsitution of Fe by Co, causing a decrease in the size of the ordered
magnetic moment. Magnetic LRO becomes unsustainable above $x=0.0125$
but the structural distortion persists until the moment in the IM state is no longer strong
enough sustain it. The IM state may be
characterized by a long, but finite magnetic correlation length and
nematic order, which biases the magnetic fluctuations along stripes.
Competing directly with magnetism is superconductivity which, on reaching full
volume fraction, weakens the magnetism dramatically and is seen directly to depress the
magnetic and structural order parameters at low temperature.

This work is supported by EPSRC (UK). Part of this work was carried
out at the Swiss Muon Source, Paul Scherrer Institut, CH, the
STFC ISIS Facility, UK and ESRF, Grenoble, FR. We thank A.\ Coldea, H.\ Luetkens and
I.\ Mazin for useful discussions and Adrian Hill (ESRF) for technical assistance on ID31.

\end{document}